\documentclass[letter,twocolumn]{jpsj3}
\usepackage{txfonts}

\usepackage{amsmath,amssymb}
\usepackage[usenames]{color}
\usepackage[dvipdfmx]{graphicx}
\usepackage{soul}

\title{
  Spin transport in the Quantum Spin Liquid State in the $S=1$ Kitaev model:
  role of the fractionalized quasiparticles
}

\author{Akihisa Koga$^1$, Tetsuya Minakawa$^1$, Yuta Murakami$^1$, and Joji Nasu$^2$}

\inst{
  $^1$Department of Physics, Tokyo Institute of Technology,
  Meguro, Tokyo 152- 8551, Japan\\
  $^2$Department of Physics, Yokohama National University,
  Hodogaya, Yokohama 240-8501, Japan
}

\abst{
  We investigate the real-time spin response of the $S=1$ Kitaev model
  upon stimuli of a pulsed magnetic field in the one of the edges
  using the exact diagonalization method.
  It is found that the pulsed magnetic field
  has no effect on the appearance of the spin moments in the quantum spin liquid region, but
  induces the spin oscillations in the other edge region
  with a small magnetic field.
  This is understood by the existence of the itinerant quasiparticles,
  which carries the spin excitations without the spin polarization
  in the quantum spin liquid state.
  This suggests that the spin fractionalizations occur
  in the $S=1$ Kitaev model as well as the exactly solvable $S=1/2$ Kitaev one
  and the fractionalized quasiparticles play an essential role
  in the spin transport.
}

\begin{document}

\maketitle

The Kitaev model has attracted much interest
since the proposal of the quantum spin model by A.~Kitaev~\cite{Kitaev2006} and
suggestion of its implementation in real materials~\cite{Jackeli2009}.
This model is composed of direction-dependent Ising exchange interactions on a honeycomb lattice, which is exactly solvable and its ground state is a quantum spin liquid (QSL)
with short-range spin correlations.
In this model, quantum spins are fractionalized into
the localized and itinerant Majorana fermions due to the quantum many-body
effect~\cite{Kitaev2006,frac1,frac2,frac3}.
The Majorana fermions have been observed recently
as a half-quantized plateau in the thermal quantum Hall experiments
in the candidate material $\alpha$-$\rm RuCl_3$~\cite{Plumb,Kubota,Sears,Majumder,Kasahara}.
Furthermore, it is theoretically clarified that
distinct energy scales ascribed to the fractionalization appear in the thermodynamic properties
such as a double-peak structure in the specific heat~\cite{Nasu1,Nasu2},
which stimulates further theoretical and experimental investigations
on the spin fractionalization~\cite{Chaloupka_2010,Yamaji_2014,Katukuri_2014,Suzuki_2015,Yamaji_2016,Kato_2017}.
Recently, the generalization of the Kitaev model with arbitrary spins~\cite{Baskaran}
has been studied,~\cite{Suzuki_2017,S1Koga,Oitmaa,Minakawa1,MixedKoga,Stavropoulos,Lee,Dong,Zhu,Khait}
where similar double peaks in the specific heat have been reported~\cite{S1Koga}.
Therefore, the spin fractionalizations are naively expected
even in the spin-$S$ Kitaev model although it is no longer solvable.

In our previous manuscipt~\cite{Minakawa2},
we have studied the real-time dynamics of
the $S=1/2$ Kitaev model by means of the Majorana mean-field theory.
It has been found that, even in the Kitaev QSL with extremely short-ranged spin correlations, 
the spin excitation propagates in the bulk without spin polarization.
This suggests that the spin transport is not caused by the change of local magnetization,
but is mediated by the itinerant Majorana fermions.
Therefore, the real-time simulation for the spin transport
is one of the promising approaches to examine the existence of
the itinerant quasiparticles in the spin-$S$ Kitaev model.

In this paper, we investigate the real-time dynamics of the $S=1$ Kitaev model
on the honeycomb lattice with two edges by means of the exact diagonalization method.
We demonstrate that after the pulsed magnetic field is applied to one of the edges,
the oscillation of spin moments does not appear in the bulk,
but is induced in the other edge region under the small magnetic field.
These results are essentially the same as those in the $S=1/2$ Kitaev model
discussed in our previous paper~\cite{Minakawa2}.
Therefore, our results support the existence of the spin fractionalization 
in the $S=1$ Kitaev model.

\begin{figure}[htb]
 \centering
 \includegraphics[width=\linewidth]{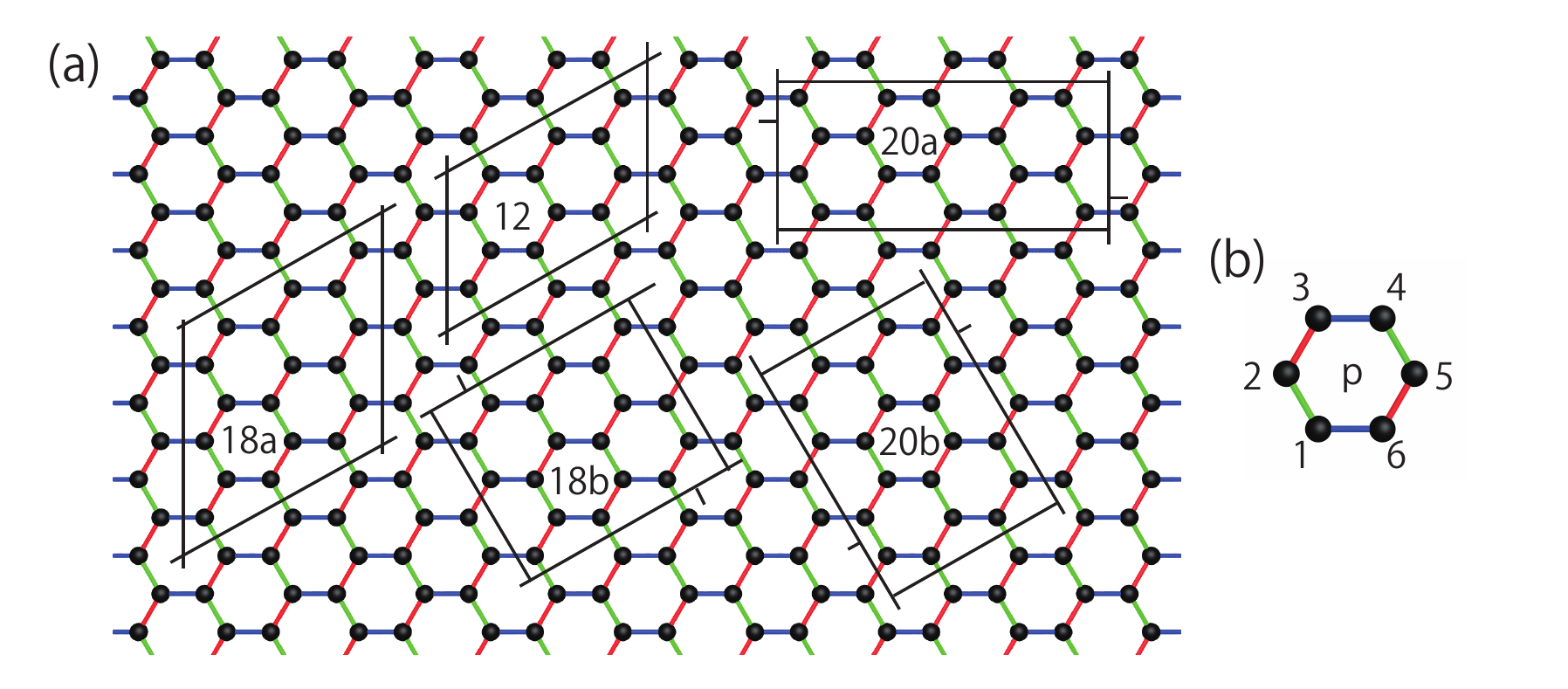}
 \caption{
   (a) $S=1$ Kitaev model on the honeycomb lattice.
   Red, blue, and green lines represent the $x$, $y$, and $z$ bonds,
   respectively.
   Clusters used in the exact diagonalizations are presented by
   the black lines.
   (b) Plaquette with sites marked 1-6 shown for the local operator
   $W_p$.
 }
 \label{fig:model}
\end{figure}

Let us consider the $S=1$ Kitaev model on the honeycomb lattice,
which should be described by the Hamiltonian as
\begin{align}
  \mathcal{H}&=\mathcal{H}_0+\mathcal{H}',\label{H}\\
  \mathcal{H}_0&=-J\sum_{\langle i,j \rangle_x}S_i^xS_j^x
  -J\sum_{\langle i,j \rangle_y}S_i^yS_j^y
  -J\sum_{\langle i,j \rangle_z}S_i^zS_j^z,\\
  \mathcal{H}'&=-\sum_ih_iS_i^z,
\end{align}
where $S_i^\gamma$ is a $\gamma(=x,y,z)$ component of an $S=1$ spin operator
at the $i$th site in the honeycomb lattice.
$J$ is the ferromagnetic exchange between the nearest-neighbor spin pairs
$\langle ij\rangle_\gamma$ on the $\gamma$-bond, and 
$h_i$ is the site-dependent magnetic field applied in the $z$-direction.
The model is schematically shown in Fig.~\ref{fig:model}(a).

We note that the Hamiltonian Eq.~(\ref{H}) has a parity symmetry.
This is clearly found if one considers
the conventional local basis sets $|m\rangle$ with $m=-1,0,1$,
which are the eigenstates of the $S^z$.
The interactions $S_i^xS_j^x$ and $S_i^yS_j^y$ inclement or declement
of both $m_i$ and $m_j$, while $S_i^zS_j^z$ and $S_i^z$ do not change them.
Therefore, the Hamiltonian Eq.~(\ref{H}) has
a global parity symmetry for $S_{tot}^z(=\sum_i S_i^z)$.
In other words, the operator
$P_z=\exp\left[ i\pi S_{tot}^z\right]$
commutes with the Hamiltonian.
This leads to the absence of the magnetization in $y$ and $z$ directions for any sites,
$\langle S_i^x\rangle=\langle S_i^y\rangle=0$
since $S_i^x$ and $S_i^y$ change the parity.
Then, in the system, the magnetization appears
only in the $z$-direction.

When no mangetic field is applied ($h_i=0$),
the ground-state and finite-temperature properties
in the $S=1$ Kitaev model has been discussed
so far~\cite{Baskaran,Suzuki_2017,S1Koga,Oitmaa,MixedKoga,Minakawa1,Stavropoulos,Lee,Dong,Zhu,Khait}.
In the case, the Kitaev model has the local $Z_2$ symmetry
on each plaquette.
The corresponding operator~\cite{Baskaran,MixedKoga} is given as,
\begin{eqnarray}
  W_p=\exp\left[i\pi \left(S_1^x+S_2^y+S_3^z+S_4^x+S_5^y+S_6^z\right)\right],\label{eq:Wp}
\end{eqnarray}
where site indexes $1,2, \cdots,6$ are introduced for a plaquette $p$,
as shown in Fig.~\ref{fig:model}(b).
This operator satisfies $W_p^2=1$ and $[\mathcal{H}_0, W_p]=0$.
Furthermore, $W_p$ commutes with $W_q$ on any plaquettes $q$.
Therefore,
the Hilbert space of the Hamiltonian $\mathcal{H}_0$ can be classified into
each subspace ${\cal S}[\{w_p\}]$ specified by the set of $w_p(=\pm 1)$, which is the eigenvalue of $W_p$.
When a state is in a certain subspace as
$|\psi\rangle=|\psi;\{w_p\}\rangle$,
the expectation value of the spin operator at the $i$th site vanishes as
$\langle \psi|S_i^\gamma|\psi \rangle=0$.
This can be proved when there exists a plaquette satisfying
the anticommutation relation $\{S_i^\gamma, W_p\}=0$.
In fact, $\langle \psi|\{S_i^\gamma, W_p\}|\psi\rangle=
\langle \psi|S_i^\gamma W_p|\psi\rangle + \langle \psi|W_pS_i^\gamma|\psi\rangle
=2w_p\langle \psi|S_i^\gamma|\psi\rangle=0$.
Furthermore, examining $\langle \psi|\{S_i^\gamma, W_p\}S_j^\gamma|\psi\rangle$,
one obtains $\langle \psi|S_i^\gamma S_j^\gamma|\psi\rangle=0$
except for the case with sites $i$ and $j$ located on the same $\gamma$ bond.
Therefore, the existence of the local conserved quantity
guarantees that the ground state of the $S=1$ Kitaev model with $h=0$
is the quantum spin liquid state
with extremely short-ranged spin-spin correlations~\cite{Baskaran}.

This discussion may be applicable in the original Hamiltonian $\mathcal{H}$
with nonuniform magnetic field $h_i$.
When no magnetic field is applied to $i(=1, 2, 4, 5)$th sites
for a plaquette $p$ in Fig.~\ref{fig:model}(b), $[\mathcal{H}, W_p]=0$.
Then, the Hilbert space is classified by eigenvalues $w_p$ in ${\cal P}$,
where ${\cal P}$ stands for the set of the plaquettes satisfying such commutation relations.
Then, we obtain $\langle S_i^z\rangle=0$ at the $i(=1, 2, 4, 5)$th sites
in the plaquette $p$ in ${\cal P}$ since $\{S_i^z, W_p\}=0$.
On the other hand, as for the plaquettes not belonging to ${\cal P}$,
the corresponding operators do not commute with the Hamiltonian
due to the presence of the magnetic field and
one cannot prove the absence of the spin moments at the site 1, 2, 4, and 5.
If the number of plaquettes not belonging to ${\cal P}$ is nonzero,
the correlation function between corresponding spins
can be finite in general.
This results from the lack of the local $Z_2$ symmetry in the Kitaev model, and thereby
it is highly nontrivial whether or not correlations indeed exist
between spins, in particular,
even when these spins are separated by the quantum spin liquid region
with extremely short-ranged spin-spin correlations.

\begin{figure}[htb]
 \centering
 \includegraphics[width=\linewidth]{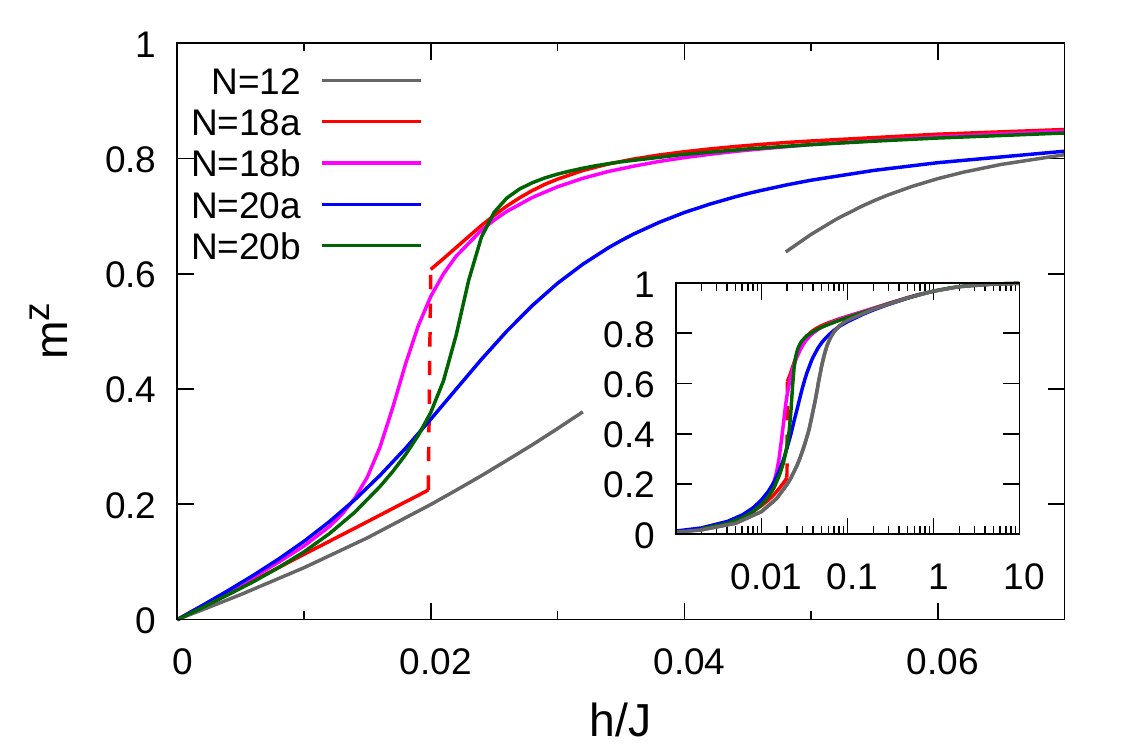}
 \caption{
   Magnetization process in the $S=1$ Kitaev systems
   with $N=12, 18a, 18b, 20a$, and $20b$.
   The vertical dashed line represents the jump singularity of $m^z$ in the $18a$ cluster.
 }
 \label{magpro}
\end{figure}

\begin{figure}[htb]
 \centering
 \includegraphics[width=\linewidth]{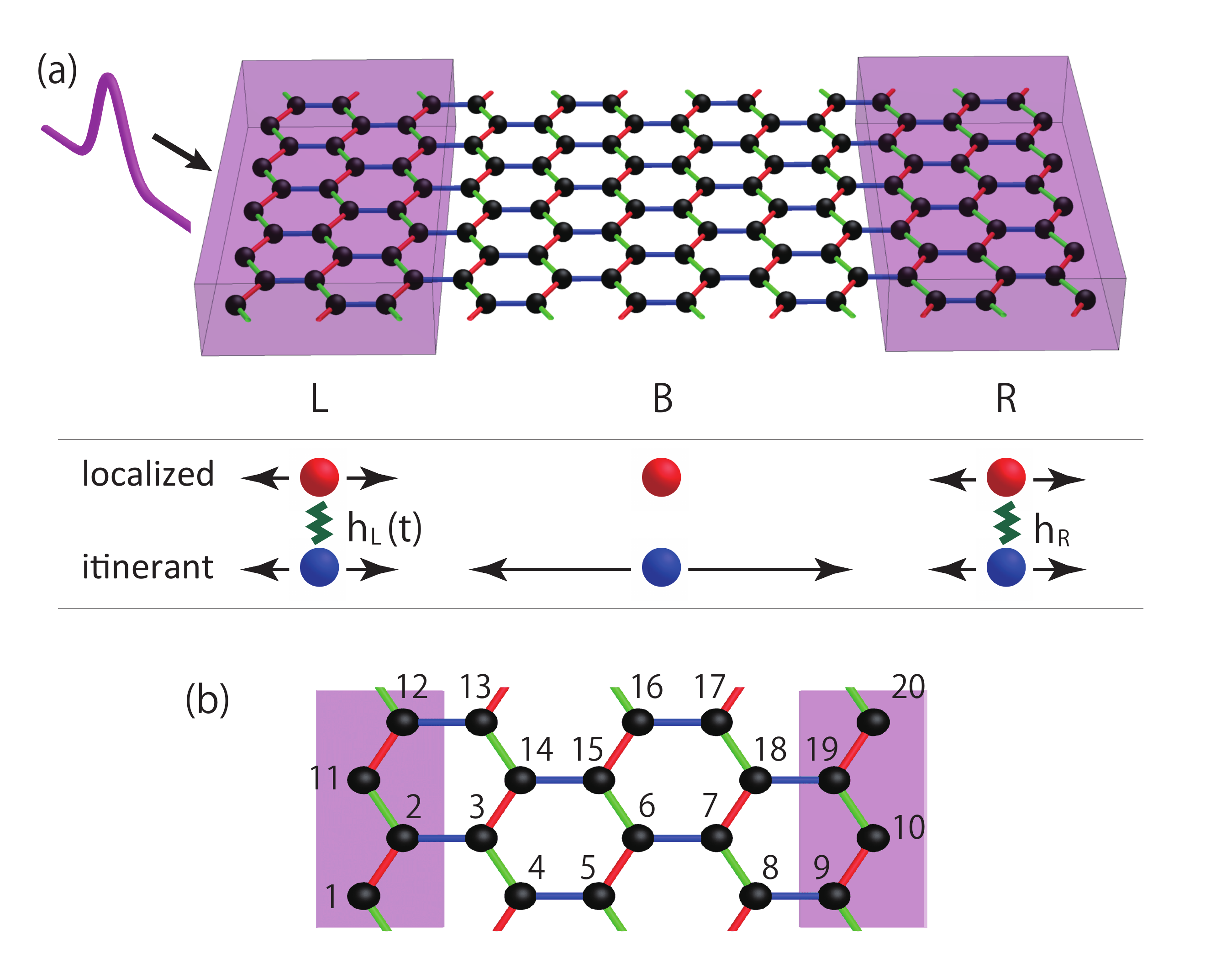}
 \caption{
   (a) Schematic picture of the Kitaev system
   on the honeycomb lattice with two edges.
   The lattice is composed of three regions.
   The static magnetic field $h_R$ is applied in the right (R) region,
   where the magnetization appears.
   In the bulk (B) region, no magnetic field is applied and
   the QSL state is realized without the magnetization.
   Time-dependent pulsed magnetic field is introduced in the left (L) region.
   (b) 20-site cluster used in the exact diagonalization.
   The numbers represent the index of the lattice site.
 }
 \label{junc}
\end{figure}

In our paper, to discuss spin-spin correlations in the Kitaev model,
we examine spin transport in the system, where
the QSL region is present between the regions
under the magnetic field [see Fig.~\ref{junc}(a)].
Before showing the results,
we briefly examine how the uniform magnetic field $h(=h_i)$ affects
the $S=1$ Kitaev model in the bulk.
By making use of the exact diagonalization for some clusters
with the periodic boundary condition (see Fig.~\ref{fig:model}),
we calculate the magnetization $m_i^z(=\langle S_i^z\rangle)$,
as shown in Fig.~\ref{magpro}.
When $h=0$, the QSL ground state is realized with $m^z=0$.
Switching on the magnetic field,
the magnetic moment is immediately induced, as shown in Fig.~\ref{magpro}.
Around $h_c/J\sim 0.02$, the magnetization rapidly increases.
In the region, the large system size dependence is observed.
This appears to be consistent with the results with $h_c\sim 0.01J$
in the (111)-direction magnetic field~\cite{Lee,Zhu,Khait}.
On the other hand, we have confirmed that the ground state always
belongs to the subspace with even parity,
including the results in the $N=18a$ cluster with a jump singularity in the magnetization process.
Therefore, it is still unclear whether or not the phase transition occurs
to the polarized state in the thermodynamic limit.
We also note that these results for the $S=1$ system are similar to
those in the $S=1/2$ Kitaev model,
where the phase transition occurs to the polarized state
at $h/J\sim 0.042$ within the mean-field theory~\cite{NasuMF}.
Therefore, we believe that there exists the energy scale characteristic of
the spin excitations in the $S=1$ Kitaev model.
In the following, we deal with the system
with a tiny magnetic field $(<h_c)$
to discuss the existence of the spin fractionalization
in the $S=1$ Kitaev model.


To study the spin transport in the $S=1$ Kitaev model,
we consider the site-dependent magnetic field,
which is defined as
\begin{equation}
  h_i=\left\{
  \begin{array}{ll}
    h_L(t)& i\in \textrm{L}\\
    0 & i\in \textrm{B}\\
    h_R&i\in \textrm{R}
  \end{array}
  \right.,
\end{equation}
where $h_L (h_R)$ is the time-dependent (static) magnetic field
applied to the left (L) [right (R)] region.
In the bulk (B) region, no magnetic field is applied,
and the QSL state is always realized without the magnetization.
The system is schematically shown in Fig.~\ref{junc}(a).

We explain the outline of our real-time simulations
by means of the exact diagonalization.
The initial ground state $|\psi\rangle$
at $t\rightarrow -\infty$ is obtained
by means of the Lanczos and inverse iteration methods.
The time-evolution of the wave function is calculated
by the time-dependent Schr\"odinger equation as,
\begin{eqnarray}
  i\frac{d}{dt}|\psi(t)\rangle=\mathcal{H}(t)|\psi(t)\rangle.
\end{eqnarray}
Then, we compute the magnetization and
nearest-neighbor spin-spin correlation on the $\gamma$-bond,
which are given as
\begin{eqnarray}
  m_i^z(t)&=&\langle \psi(t)|S_i^z|\psi(t)\rangle,\\
  C_{ij}(t)&=&\langle \psi(t)|S_i^\gamma S_j^\gamma|\psi(t)\rangle.
\end{eqnarray}
In this study, we introduce the pulsed magnetic field in the L region,
which is explicitly given as
\begin{eqnarray}
  h_L(t)=\frac{A}{\sqrt{2\pi\sigma^2}}\exp\left[ -\frac{t^2}{2\sigma^2}\right],
\end{eqnarray}
where $A$ and $\sigma$ are magnitude and width of the Gaussian pulse.
In the following, we fix $A=1$ and $\sigma=2/J$.
\begin{figure}[htb]
 \centering
 \includegraphics[width=\linewidth]{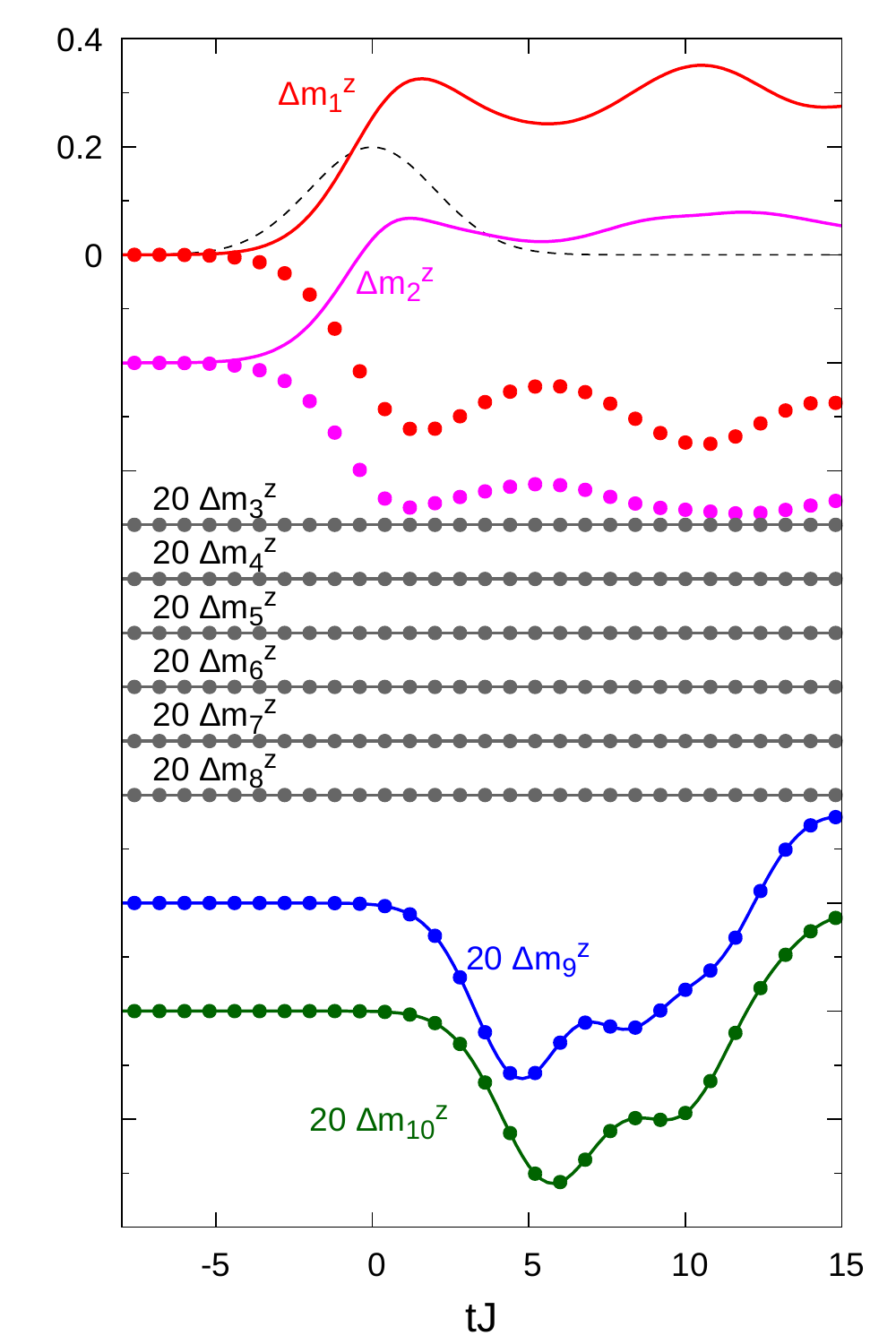}
 \caption{
   Real-time evolution of the change in the local magnetization
   in the system with $h_R/J=0.01$
   after the introduction of the pulsed magnetic field
   with $A=1$ and $\sigma=2/J$ shown as the dashed line (offset for clarity).
   Dotted lines represent the results for the pulse
   with $A=-1$ and $\sigma=2/J$.
 }
 \label{mag_t}
\end{figure}

In this calculation, we examine the 20-site cluster with two edges, where the periodic boundary condition is imposed for the direction perpendicular to the edge,
as shown in Fig.~\ref{junc}(b).
In the cluster, the R and L regions include four sites.
There exist twelve sites in the B region, where
no magnetic field is applied.
Although the cluster we treat is too small,
the spin transport characteristic of the Kitaev system is expected to be captured
since there exist four plaquettes with the local $Z_2$ symmetry, which is crucial for the peculiar spin transport, in the B region.
In fact, we have confirmed that, in the initial state $(t\rightarrow -\infty)$, 
the site-dependent magnetization $m_i^z$ appears
only in the R region $(m_9^z=m_{19}^z=0.217$ and $m_{10}^z=m_{20}^z=0.220$).
Figure~\ref{mag_t} shows the time dependence of the change in the spin moment
$\Delta m^z_i[=m_i^z(t)-m_i^z(-\infty)]$
after the pulsed magnetic field is introduced.
The magnetic moments at the sites 1 and 2 in the L region are induced
by the pulsed magnetic field.
On the other hand, no magnetic moment appears in the B region,
which is consistent with the existence of the local conserved quantity.
On the other hand, after some delay, the spin oscillation is induced
at the sites 9 and 10 in the R region.
This means that the wave-packet triggered by the magnetic pulse
in the L region reaches the R region
through the B region without spin oscillations.
Since $m_i^x(t)=m_i^y(t)=0$ for any sites,
we can say that the spin moment plays no role in the spin transport in the $S=1$ Kitaev model.
This remarkable phenomenon is similar to that in the $S=1/2$ Kitaev model~\cite{Minakawa2},
where the spin transport is mediated by the itinerant Majorana fermions.
Therefore, our results suggest the existence of the itinerant quasiparticles,
which is not associated with the spin polarization even in the $S=1$ Kitaev model.
Namely, we expect that
in the $S=1$ Kitaev model without a static magnetic field,
the spin degree of freedom is fractionalized 
into two; itinerant and localized quasiparticles owing to the existence of the local $Z_2$ symmetry.
The pulsed magnetic field in the L region creates
the itinerant and localized quasiparticle excitations,
while only the formers propagate in the whole system.
In the R region,
the itinerant and localized quasiparticles are hybridized by the static magnetic field
due to the lack of the local $Z_2$ symmetry, leading to the finite spin oscillations.
The scenario for the spin transport 
is schematically shown in Fig.~\ref{junc}(a).

The spin fractionalization in the $S=1$ Kitaev model has been suggested
in the thermodynamic properties such as a double-peak structure
in the specific heat~\cite{S1Koga}.
We note that the higher temperature peak is closely related to
the nearest-neighbor spin-spin correlations $C_{ij}$.
It is known that the higher temperature peak in the $S=1/2$ case corresponds to
the motion of the itinerant Majorana fermions.
Therefore, one can expect that the flow of $C_{ij}$ is regarded as the motion of
the itinerant quasiparticles in the $S=1$ Kitaev model. 
\begin{figure}[htb]
 \centering
 \includegraphics[width=\linewidth]{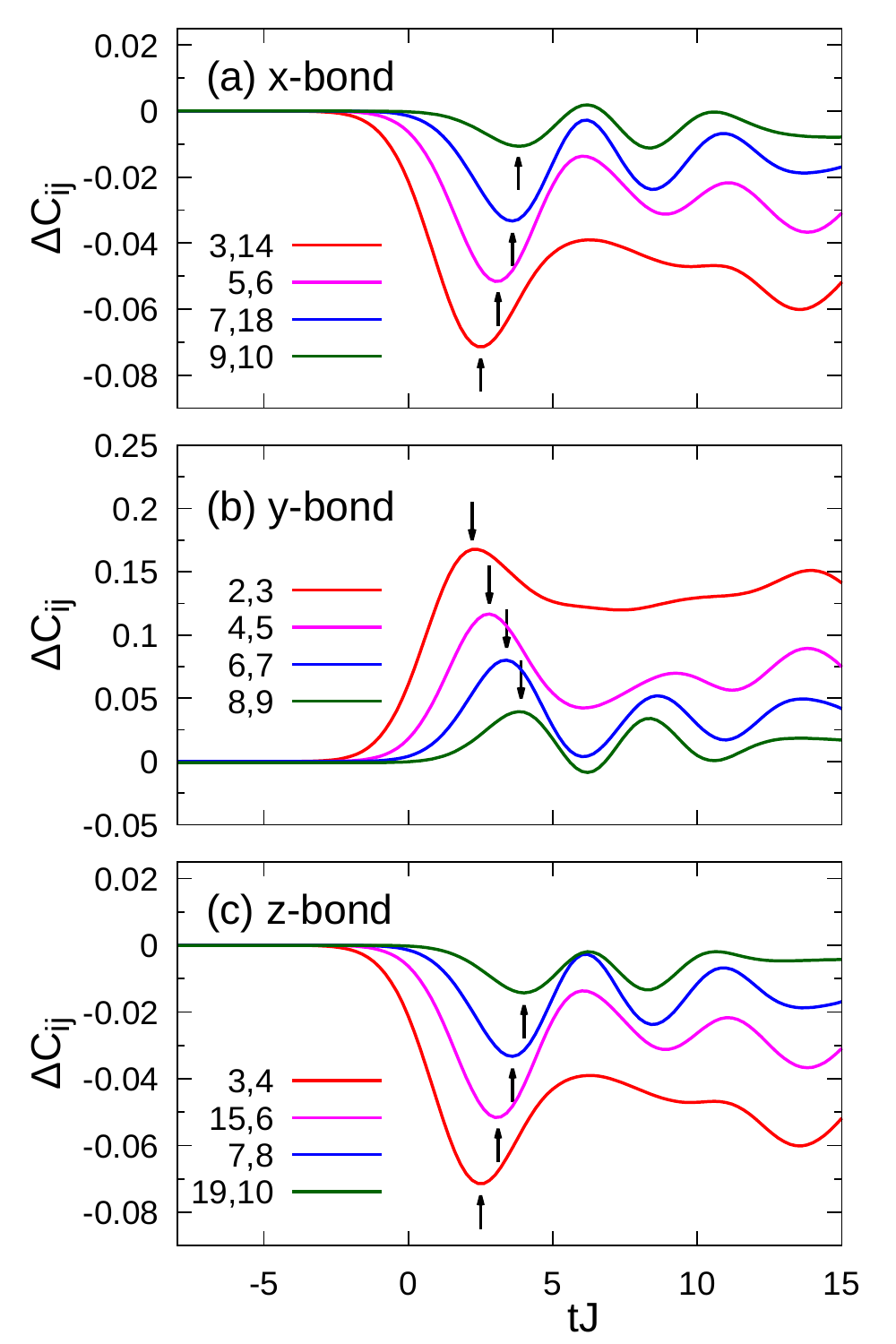}
 \caption{
   Real-time evolution of the change in the nearest-neighbor correlations
   on the (a) $x$-bond, (b) $y$-bond, and (c) $z$-bond
   in the B and R region of the $S=1$ Kitaev system with $h_R/J=0.01$, $A=1$, and $\sigma=2/J$.
   Pairs of the numbers indicate two sites coupled by the Kitaev exchanges
   in the 20-site cluster shown in Fig.~\ref{junc}.
 }
 \label{fig:corr}
\end{figure}
Figures~\ref{fig:corr}(a), \ref{fig:corr}(b) and \ref{fig:corr}(c) show the real-time evolution of 
the change in the nearest-neighbor spin correlations on the $x$-, $y$-, and $z$-bonds
except for the L region.
Oscillations appear in the spin-spin correlations on all exchanges
even in the B region.
We also find that the change of the quantities becomes small with increasing the distance from the L region.
Associated with this change, the characteristic time,
where $|\Delta C_{ij}|$ takes a first maximal value
(shown as arrows in Fig.~\ref{fig:corr}),
becomes longer.
This implies that the wave-packet created by the pulsed magnetic field
at the L region indeed flows to the R region through the B region.
The second maximal values are considered to be caused by
the reflection of the flow at the right edge
since the peaks shift to the left side as time elapses.

We also note the interesting pulse-field dependence of the phenomena
originated from the $Z_2$ symmetry~\cite{Minakawa2}. 
The system has the local $Z_2$ symmetry in the L region before 
the pulsed magnetic field is applied.
In this case, each eigenstate is specified by a set of the eigenvalues of $W_p$
in the L region.
The Hamiltonian for the magnetic pulse has the offdiagonal elements between distinct subspaces.
An important point is that
the operator $S_i^z$, in general, flips two eigenvalues of $W_p$ for adjacent plaquettes
sharing the same $z$-bond, which connects the $i$th site and its pair site.
Therefore, if the ground state belongs to the subspace with the configuration $\{w_p\}$,
only even-order perturbations in the pulsed magnetic field
contribute the expectation value for the operator $O$ satisfying $[O, W_p]=0$ with $p\in L$.
This means that this expectation value is independent of the sign of the pulsed field.
To confirm this, we calculate the time-dependent spin moments after the pulsed magnetic field
in the $-z$ direction ($A=-1$).
The results are shown as the dotted lines in Fig.~\ref{mag_t}.
In the L region, the magnetic moments are induced in the direction of the applied field.
By contrast, in the B and R regions, the results do not depend on
the sign of the pulsed magnetic field.

Finally, we briefly comment on the nature of the low-lying excitation
in the $S=1$ Kitaev system~\cite{S1Koga,Lee,Dong,Zhu,Khait}.
The real-time simulation, in principle, allows us to
clarify if the system is gapped or gapless,
by examining the velocity and decay rate of the wave packet created by
the pulsed field with small $A$ and/or large $\sigma$.
However, in the small-size numerical calculations, 
it should be hard to evaluate them 
due to the interference effect for the multiple reflections.
Therefore, further numerical calculations for the larger systems
are necessary to clarify the elementary excitations in the $S=1$ Kitaev model,
which is now under consideration.

In summary, we have studied the real-time dynamics of the $S=1$ Kitaev model
on the honeycomb lattice.
Applying the pulsed magnetic field to one of the edges in the system,
spin oscillations never appear in the bulk while
they appear in the other edges.
Similar behavior appears in the $S=1/2$ Kitaev model,
where fractionalized Majorana fermions flow in the system.
Therefore, 
our results suggest the existence of the spin fractionalization in the $S=1$ Kitaev model
and the spin transport is mediated by the fractionalized itinerant quasiparticles.
This behavior should be common in the spin-$S$ Kitaev model, which 
is consistent with thermodynamic properties
such as the double-peak structure in the specific heat
and the half-plateau in the entropy~\cite{S1Koga}.
It is also important to clarify the spin transport in the spin-$S$ Kitaev models
while finite temperature calculations suggest that the entropy of the quantum spins
is split in half into those of the itinerant and localized quasiparticles.
These interesting problems remain as future issues.

\begin{acknowledgments}
  Parts of the numerical calculations are performed
  in the supercomputing systems in ISSP, the University of Tokyo.
  This work was supported by Grant-in-Aid for Scientific Research from
  JSPS, KAKENHI Grant Nos.
  JP19K23425 (Y.M.),
  JP19H05821, JP18K04678, JP17K05536 (A.K.),
  JP16H02206, JP18H04223, JP19K03742 (J.N.),
  by JST CREST (JPMJCR1901) (Y.M.),
  and by JST PREST (JPMJPR19L5) (J.N.).
\end{acknowledgments}

\bibliographystyle{jpsj}
\bibliography{./refs}

\begin{thebibliography}{10}

\bibitem{Kitaev2006}
A.~Kitaev: Ann. Phys. (N. Y.) {\bfseries 321} (2006) 2.

\bibitem{Jackeli2009}
G.~Jackeli and G.~Khaliullin: Phys. Rev. Lett. {\bfseries 102} (2009) 017205.

\bibitem{frac1}
X.-Y. Feng, G.-M. Zhang, and T.~Xiang: Phys. Rev. Lett. {\bfseries 98} (2007)
  087204.

\bibitem{frac2}
H.-D. Chen and J.~Hu: Phys. Rev. B {\bfseries 76} (2007) 193101.

\bibitem{frac3}
H.-D. Chen and Z.~Nussinov: J. Phys. A: Math. Theor. {\bfseries 41} (2008)
  075001.

\bibitem{Plumb}
K.~W. Plumb, J.~P. Clancy, L.~J. Sandilands, V.~V. Shankar, Y.~F. Hu, K.~S.
  Burch, H.-Y. Kee, and Y.-J. Kim: Phys. Rev. B {\bfseries 90} (2014) 041112.

\bibitem{Kubota}
Y.~Kubota, H.~Tanaka, T.~Ono, Y.~Narumi, and K.~Kindo: Phys. Rev. B {\bfseries
  91} (2015) 094422.

\bibitem{Sears}
J.~A. Sears, M.~Songvilay, K.~W. Plumb, J.~P. Clancy, Y.~Qiu, Y.~Zhao,
  D.~Parshall, and Y.-J. Kim: Phys. Rev. B {\bfseries 91} (2015) 144420.

\bibitem{Majumder}
M.~Majumder, M.~Schmidt, H.~Rosner, A.~A. Tsirlin, H.~Yasuoka, and M.~Baenitz:
  Phys. Rev. B {\bfseries 91} (2015) 180401.

\bibitem{Kasahara}
Y.~Kasahara, T.~Ohnishi, Y.~Mizukami, O.~Tanaka, S.~Ma, K.~Sugii, N.~Kurita,
  H.~Tanaka, J.~Nasu, Y.~Motome, T.~Shibauchi, and Y.~Matsuda: Nature
  {\bfseries 559} (2018) 227.

\bibitem{Nasu1}
J.~Nasu, M.~Udagawa, and Y.~Motome: Phys. Rev. B {\bfseries 92} (2015) 115122.

\bibitem{Nasu2}
J.~Nasu, J.~Yoshitake, and Y.~Motome: Phys. Rev. Lett. {\bfseries 119} (2017)
  127204.

\bibitem{Chaloupka_2010}
J.~Chaloupka, G.~Jackeli, and G.~Khaliullin: Phys. Rev. Lett. {\bfseries 105}
  (2010) 027204.

\bibitem{Yamaji_2014}
Y.~Yamaji, Y.~Nomura, M.~Kurita, R.~Arita, and M.~Imada: Phys. Rev. Lett.
  {\bfseries 113} (2014) 107201.

\bibitem{Katukuri_2014}
V.~M. Katukuri, S.~Nishimoto, V.~Yushankhai, A.~Stoyanova, H.~Kandpal, S.~Choi,
  R.~Coldea, I.~Rousochatzakis, L.~Hozoi, and J.~v.~d. Brink: New J. Phys.
  {\bfseries 16} (2014) 013056.

\bibitem{Suzuki_2015}
T.~Suzuki, T.~Yamada, Y.~Yamaji, and S.-i. Suga: Phys. Rev. B {\bfseries 92}
  (2015) 184411.

\bibitem{Yamaji_2016}
Y.~Yamaji, T.~Suzuki, T.~Yamada, S.-i. Suga, N.~Kawashima, and M.~Imada: Phys.
  Rev. B {\bfseries 93} (2016) 174425.

\bibitem{Kato_2017}
Y.~Kato, Y.~Kamiya, J.~Nasu, and Y.~Motome: Phys. Rev. B {\bfseries 96} (2017)
  174409.

\bibitem{Baskaran}
G.~Baskaran, D.~Sen, and R.~Shankar: Phys. Rev. B {\bfseries 78} (2008) 115116.

\bibitem{Suzuki_2017}
T.~Suzuki and Y.~Yamaji: Phys. B: Condense. Matter {\bfseries 536} (2017) 637.

\bibitem{S1Koga}
A.~Koga, H.~Tomishige, and J.~Nasu: J. Phys. Soc. Jpn. {\bfseries 87} (2018)
  063703.

\bibitem{Oitmaa}
J.~Oitmaa, A.~Koga, and R.~R.~P. Singh: Phys. Rev. B {\bfseries 98} (2018)
  214404.

\bibitem{Minakawa1}
T.~Minakawa, J.~Nasu, and A.~Koga: Phys. Rev. B {\bfseries 99} (2019) 104408.

\bibitem{MixedKoga}
A.~Koga and J.~Nasu: Phys. Rev. B {\bfseries 100} (2019) 100404.

\bibitem{Stavropoulos}
P.~P. Stavropoulos, D.~Pereira, and H.-Y. Kee: Phys. Rev. Lett. {\bfseries 123}
  (2019) 037203.

\bibitem{Lee}
H.-Y. Lee, N.~Kawashima, and Y.~B. Kim: arXiv p. 1911.07714.

\bibitem{Dong}
X.-Y. Dong and D.~N. Sheng: arXiv p. 1911.12854.

\bibitem{Zhu}
Z.~Zhu, Z.-Y. Weng, and D.~N. Sheng: arXiv p. 2001.05054.

\bibitem{Khait}
I.~Khait, P.~P. Stavropoulous, H.-Y. Kee, and Y.~B. Kim: arXiv p. 2001.06000.

\bibitem{Minakawa2}
T.~Minakawa, Y.~Murakami, A.~Koga, and J.~Nasu: arXiv  (2019) 1912.10599.

\bibitem{NasuMF}
J.~Nasu, Y.~Kato, Y.~Kamiya, and Y.~Motome: Phys. Rev. B {\bfseries 98} (2018)
  060416.

\end{thebibliography}

\end{document}